# Effect of extended defects on AlGaN QDs for electron-pumped UV-emitters

Jesus Cañas[*,1], Névine Rochat[2], Adeline Grenier[2], Audrey Jannaud[2], Zineb Saghi[2], Jean-Luc Rouviere[3], Edith Bellet-Amalric[1], Anjali Harikumar[1], Catherine Bougerol[4], Lorenzo Rigutti[5] and Eva Monroy[1]

[1] Univ. Grenoble-Alpes, CEA, Grenoble INP, IRIG, PHELIQS, 38000 Grenoble, France
[2] Univ. Grenoble Alpes, CEA, LETI, 38000 Grenoble, France
[3] Univ. Grenoble Alpes, CEA, IRIG, MEM, 38000 Grenoble, France
[4] Univ. Grenoble Alpes, CNRS, Grenoble INP, Institut Neel, 38000 Grenoble, France
[5] UNIROUEN, CNRS, Groupe de Physique des Matériaux, Normandie Université, 76000 Rouen, France

*Corresponding author: jesus.canasfernandez@cea.fr

OrcID:
Jesus Cañas: 0000-0003-0202-6987
Névine Rochat: 0000-0003-3574-4424
Zineb Saghi: 0000-0002-8895-8410
Jean-Luc Rouviere: 0000-0001-8731-3074
Edith Bellet-Amarlic: 0000-0003-2977-1725
Anjali Harikumar: 0000-0002-0354-2142
Catherine Bougerol: 0000-0002-4823-0919
Lorenzo Rigutti: 0000-0001-9141-7706
Eva Monroy: 0000-0001-5481-3267

**ABSTRACT:** We study the origin of bimodal emission in AlGaN/AlN QD superlattices displaying high internal quantum efficiency (around 50%) in the 230-300 nm spectral range. The secondary emission at longer wavelengths is linked to the presence of cone-like defects starting at the first AlN buffer/superlattice interface and propagating vertically. These defects are associated with a dislocation that produces strong shear strain, which favors the formation of 30° faceted pits. The cone-like structures present Ga enrichment at the boundary facets and larger QDs within the defect. The bimodality is attributed to the differing dot size/composition within the defects and at the defect boundaries, which is confirmed by the correlation of microscopy results and Schrödinger-Poisson calculations.

# 1. INTRODUCTION

There is a strong interest in the development of efficient solid-state UV-C sources in the 220-270 nm range due to their powerful germicidal properties.[1–4] Light emitting diodes (LEDs) based on the AlGaN semiconductor system are the most promising approach, but the early stage of development of the AlGaN technology still presents challenges, such as inefficient p-type doping and resistive ohmic contacts, which result in low efficiencies.[5,6] An approach that circumvents these material limitations is the use of electron beam pumped lamps.[7,8] These lamps consist of a UV-emitting material (anode) housed in a vacuum tube. Within this tube, high-energy electrons are emitted from a cold cathode and accelerated towards the semiconductor anode, generating free carriers. Unlike LEDs, electron beam pumped lamps do not require p-type doping or contacts, as carrier generation occurs through the impact ionization process in a large active volume (active region depth > 100 nm). Significant progress has been made in the development of electron-pumped devices including remarkable advances in miniaturization,[9] efficiency improvements,[10] and output power,[11] with some devices now capable of reaching watt-level outputs. To further enhance device performance, the utilization of AlGaN quantum dots (QDs) has emerged as a promising avenue. The three-dimensional (3D) carrier confinement within QDs prevents carrier diffusion towards non-radiative centers, leading to improved radiative efficiency even at high temperatures.[12]

Previous studies have demonstrated $Al_xGa_{1-x}N$/AlN QDs with internal quantum efficiency (IQE) of around 50% in the UV-C range, maintaining stable efficiency even when exposed to optical pumping power levels reaching up to 200 kW/cm$^2$.[13,14] However, samples emitting at wavelengths shorter than 270 nm present broad or bimodal spectra. This phenomenon has been reported by various researchers in planar samples containing thin quantum wells or QDs for deep-UV emission.[15–18] It has been attributed to various factors, such as emission from the cladding layer, reemission from AlGaN layers, or even the impact of the free carrier density of states in disk-like structures. Likewise, a similar problem has been identified in UV-emitting nanowires, with multiple reports confirming this trend.[19–22] In these cases, secondary emission has been linked to recombination in AlGaN capping layers, defect-related emission, and fluctuations in the Ga composition. Identifying the root causes of these spectral features is crucial for developing strategies to mitigate or reduce their presence effectively.

In this work, we investigate the origin of the bimodal emission in $Al_xGa_{1-x}N$/AlN QDs superlattices designed for electron beam pumped UV emitters. With this purpose, we perform a comprehensive analysis combining scanning transmission electron microscopy (STEM), atom probe tomography (APT), electron tomography, cathodoluminescence (CL) spectroscopy and 3D simulations of the electronic structure. The bimodal emission is attributed to the presence of a secondary population of quantum dots located within cone-like structural defects. These defects emerge at the initial interface between the AlN buffer layer and the QD superlattice and propagate vertically. They are associated to a dislocation featuring a strong strain field, thereby facilitating the generation of a pit characterized by 30° faceted boundaries.

## 2. EXPERIMENTAL SECTION

The samples under study consist of 100 periods of $Al_xGa_{x-1}N$/AlN self-assembled QDs deposited on commercial 1-μm-thick (0001)-oriented AlN-on-sapphire templates by PAMBE. During the growth, the substrate temperature was fixed at 720°C, and the active nitrogen flux was adjusted to provide a growth rate of 0.6 monolayers per second (ML/s) under metal-rich conditions. For the generation of $Al_xGa_{x-1}N$ QDs, the metal/N flux ratio was kept below the stoichiometric value. This resulted in the Stranski-Krastanov (SK) growth mode due to the lattice mismatch and the energetically favorable {10-13} facets over the (0001) plane, under N-rich conditions.[23] The growth process was monitored by reflection high-energy electron diffraction (RHEED). The Al$_x$Ga$_{x-1}$N deposition time was set to 15 s, during which roughening was observed in the RHEED pattern. Two samples with different aluminum concentration in the QDs are studied in this work: S1 with 14% Al content and S2 with 46% of Al. The AlN overgrowth to form the barriers was performed under Al-rich conditions (Al/N ≈ 1.1), to favor the planarization of the surface. After growing 4 nm of AlN, the Al cell was shuttered and the excess of metal was consumed with nitrogen. A schematic description of the complete structures is presented in figure 1a.

A TEM lamella and a needle-shaped specimen for atom probe and electron tomography were prepared using a scanning electron microscope/focused ion beam (SEM/FIB) Helios Nanolab 450S operated at 16 kV. Low energy (2 keV) ion beam was used during the final APT annular milling in order to reduce irradiation damages on the sidewalls of the specimen.

The structural properties of the samples were studied by bright field (BF) and high-angle annular dark field (HAADF) STEM imaging using a probe-corrected Titan Themis (Thermo Fisher Scientific) operating at 200 kV. The 3D morphology of the extended defect was investigated by electron tomography on a needle-shaped specimen mounted on a Fischione on-axis tomography holder (model 2050). The HAADF-STEM tilt series was acquired over a tilt range of 180 degrees, in 2-degree steps. The tilt series was then aligned by cross-correlation and the common-line method[24], and the 3D volume was reconstructed using the Simultaneous Iterative Reconstruction Technique (SIRT) implemented in Astra toolbox. 3D visualization was performed using TomViz platform. Nanobeam Precesion electron diffraction[25] (N-PED) experiments were carried out using the same microscope as above. Precession Electron Diffraction Maps (PED-M) were acquired using the PrecessionMkII software. In PrecessionMkII, the fast scanning unit of STEM is used to precess the beam. The scanning of the beam is performed by controlling numerically the slower beam shift coil using the internal software of the microscope. The precession speed was set to its lowest value, i.e. 0.1 ms per rotation, layers were oriented along on the [1 1 -2 0] direction and the convergence angle was set to 4 mrad using a condenser aperture of 10 micron giving a probe diameter at half maximum of about 0.3 nm. Strain was determined by measuring the distance between diffraction spots.

Three-dimensional compositional maps of the samples were measured by laser-assisted atom probe tomography using a CAMECA FlexTAP instrument. The system is equipped with a pulsed laser operating at a wavelength of 343 nm, with a pulse duration of 450 fs, an energy of 16 nJ/pulse and a repetition rated of 100 kHz repetition rate. The tip temperature was maintained around 40 K during analysis. The field of view was ±15°; roughly equivalent to a straight flight atom probe with a 74 cm flight-path. The APT experiment was performed at low electric field conditions on the tip surface, as suggested by the charge-to-state ratio $Ga^{2+}/Ga^{+}$~0.03, corresponding to a surface field of 230 MV/cm.[26] This allows to observe the best signal-to-noise ratio in the mass spectrum[27,28] and reduce the preferential loss of metallic species during the tip evaporation[29,30]. APT data were reconstructed using a standard shank angle reconstruction algorithm[31,32].

Cathodoluminescence measurements were carried out using an Attolight CL-dedicated SEM microscope. The acceleration voltage was 10 kV and the beam current was ≈ 5 nA. The luminescence was collected through an integrated microscope objective (numerical aperture = 0.7). By scanning the sample, the optical spectra of each pixel are recorded on a CCD camera through a dispersive spectrometer with 400 mm focal length (grating: 150 grooves/mm blazed at 500 nm).

The electronic structure of the QDs was modelled using the Nextnano3 8-band k·p Schrödinger-Poisson equation solver[33] with the GaN and AlN parameters described by Kandaswamy, et al.[34]. For $Al_xGa_{1-x}N$ alloys, all the bowing parameters were set to zero. We used a model structure consisting of a 10-period superlattice of QDs, with a close-packed distribution of the in-plane QDs (modelling 7 QDs per plane), and assuming that the QDs are vertically aligned. The strain distribution in the structure is calculated by minimization of the elastic energy prior to the calculation of the band diagram, which considered the spontaneous and piezoelectric polarization and the band gap deformation potentials. The electronic levels were calculated in a QD in the center of the structure. The variation of the exciton binding energy was taken into account as a one-dimensional correction.[14]

## 3. RESULTS AND DISCUSSION

This study investigates bimodal emission in AlGaN/AlN QD superlattices, primarily focusing on sample S1, with QDs containg 14% of Al and emitting at 294 nm at room temperature. The findings and discussion are organized in three sections. We start by presenting the structural analysis and morphology of AlGaN QDs in S1. Then, the cathodoluminescence of this sample is discussed and correlated with the defect structure, including experiments and calculations. Finally, in the third section, the study expands its purview to shorter wavelengths emitters by incorporating sample S2, with QDs containing 46% of Al and emitting at 233 nm.

**Structural Analysis and Morphology of the AlGaN QDs**

A BF-STEM image of the full structure of S1 is presented in figure 1b. This image provides a comprehensive visualization of the sample, including the AlN substrate, a thin AlN buffer layer (< 100 nm), and the $Al_{0.14}Ga_{0.86}N$/AlN QD superlattice. A notable density of threading dislocations is observed within the region of interest, predominantly originating from the substrate, with a dislocation density of approximately $10^9$ cm$^{-2}$. A zoomed-in HAADF-STEM image in figure 1c highlights three QD planes of a defect-free region. In the projected image, the QD layers look like rough quantum wells due to the fact that the QD diameter is smaller than the thickness of the lamella specimen (around 70 nm), which results in more than one dot being projected in the images. The QD shape can be inferred from the analysis of the in-plane contrast variations. Through detailed analysis of STEM images of about 50 QDs in homogeneous regions of the sample, we estimate the QD diameter to be 6.2 ± 1.3 nm along the $\langle 1\bar{1}00 \rangle$ direction, with a QD height around 5.0 ± 0.6 ML including the wetting layer, whose thickness is estimated at 2.5 ± 0.5 ML. Taking into consideration the preferential {10-13} facets identified in previous works,[36,37] the size and morphology of the QDs is depicted in figure 1d.

**Cathodoluminescence: Correlation with the Defect Structure**

The surface of the S1 sample unveils a high density of pits, as showcased in the SEM image presented in figure 2a. Figure 2b presents the top-view CL spectra of the S1 sample, measured at room temperature and 5 K, recorded on the dashed square region highlighted in figure 2a. The peak emission is observed at 291 nm for 5 K and at 294 nm for room temperature. By integrating the full CL spectra intensities at room temperature and 5 K and comparing their values, the $IQE$ is estimated to be 46%[14]. This value remains consistently high (46 ± 6 %) throughout the sample and aligns with our previous report of IQE around 50% for QDs emitting between 320 nm and 230 nm.[13]

If we focus on the shape of the spectrum at 5 K, it exhibits two distinct contributions: a more intense emission centered at 291 nm and a secondary, less intense emission at 302 nm. These two contributions merge in the room temperature spectrum, resulting in a wide emission peak with a full width at half maximum FWHM = 22 nm. Narrower emission lines are much preferred for disinfection, in order to target the more efficient spectral range and avoid exposure to inefficient but harmful radiation.

To analyse the spatial distribution of the two spectral contributions, we have carried out CL mapping experiments. Upon comparing the morphology of the sample with the associated λ-filtered CL map in figure 2c (colors indicated in figure 2b), it becomes evident that the emission at longer wavelengths is linked to the pits. The regions emitting at longer wavelengths are spatially localized and separated by hundreds of nanometers. They appear with a density of approximately $2\times10^8$ cm$^{-2}$. Overall, the emission spectra of the flat surface peak is around 290 nm at 5 K and 294 nm at room temperature. In contrast, the emission in the vicinity of the pits shifts to 302 nm at 5 K and to 306 nm at room temperature, as depicted in figure 2d. Surprisingly, the radiative efficiency does not decrease, as one might expect if the secondary emission originated from point defects.

In order to gain a deeper understanding of the structural and optical effects associated with the pits observed at the surface, a TEM lamella was prepared and analyzed using STEM and CL. Figure 3a presents an SEM image of the lamella extracted from the S1 sample, with a superimposed λ-filtered CL map which reveals a relatively uniform emission at the main wavelength of 290 nm. However, localized regions within the lamella exhibit emission at a longer wavelength, around 302 nm, which should correspond to the longer wavelength CL observed from the surface (figure 2b). These regions appear as cone-shaped structures, extending from the base of the superlattice to the middle or top of the sample.

Further insights were obtained from a BF-STEM image of one of the longer-wavelength emitting region, which is presented in figure 3b. In addition to the threading dislocations, we observe 3D defects with conical shape that originate at the first interface of the superlattice and propagate vertically. They cause a bending of the QD planes and barriers within the superlattice, and appear associated with the surface pits.

Figure 3c shows a HAADF-STEM image of the same region as figure 3b with a superimposed λ-filtered CL map. The extended defects are clearly associated with the longer wavelength emission, while the threading dislocations, present in higher density, do not significantly affect the wavelength or intensity of the luminescence. Figure 4a shows a HAADF-STEM image of the base of one of the extended defects, including the interface between the AlN buffer layer and the superlattice. The image reveals a threading dislocation that originates from the AlN substrate. In close vicinity with the dislocation, the AlN buffer layer exhibits a step with an inclined facet forming approximately 30° with the (0001) plane. The faceted step propagates along the superlattice.

The presence of steps can be attributed to the strong strain field in the vicinity of the dislocation, which promotes the emergence of facets with low surface energy. Indeed, Figure 4b presents deformation maps around the base of the extended defect. The $\varepsilon_{zz}$ map shows the variation of the out-of-plane lattice parameter with respect to a reference, which is the AlN lattice constant in the buffer layer. The image reveals an increased lattice parameter in the AlGaN QD planes (red/yellow), but also in the AlN barriers (green), which are slightly deformed along the growth axis due to the presence of the QDs. The $\varepsilon_{xx}$ map shows the variation of the in-plane lattice parameter with respect to the AlN in-plane lattice constant in the buffer layer. Note that, away from the dislocation, the AlGaN QD planes and AlN barriers exhibit a similar in-plane lattice parameter, which confirms that the embedded QDs introduce only an elastic deformation of the lattice. However, the lattice parameter is enlarged along the dislocation and the faceted step. Finally, the $\varepsilon_{xz}$ map illustrates the shear component of the strain, which is slightly negative in the QD layers due to the hydrostatic pressure imposed by the AlN matrix into the dots. On the contrary, the presence of the dislocation generates a strong shear strain around it, as shown by $\varepsilon_{xz}$ becoming positive along the dislocation line and extending along the faceted step. Thus, once generated, the step propagates vertically through the superlattice. The approximate angle of 30° between the (0 0 0 1) surface and the exposed facets on the steps indicates that these facets correspond to either {5 -5 0 16} or {1 0 -1 3} planes, which are known to have a low surface energy.[23,38]

The presence of the extended defect affects the morphology of the QDs. Dots outside but in the vicinity of the defective region, highlighted in figure 4c, present the same morphology and size as those in the homogeneous region depicted in figure 1c. In contrast, QDs located inside the extended defect, for example those in figure 4d, exhibit an increased height of 6 ± 0.5 ML, and their diameter expands to approximately 7.2 ± 1.2 nm. Note that their aspect ratio remains at 0.12, relatively similar to those in the homogeneous region. This implies that the elastic energy accumulated in the proximity of the defect would contribute to enhance the mobility of Ga adatoms during the QD deposition, leading to a higher capture radius around the QD nucleation site. The increased size of the QD population within the extended defects can shift the emission wavelength explaining the bimodal emission observed in the CL maps.

To gain further insight into the 3D structure of the defects, HAADF-STEM tomography was performed on a needle-shaped specimen, obtained from sample S1, containing an extended defect. A slice through the tomographic reconstruction is presented in figure 5a, showing the propagation of the pit-like structure through the superlattice. Interestingly, a chemical-related bright contrast is observed at the boundaries of the extended defect, indicating a higher Ga content compared to the surrounding regions. This higher Ga concentration can partially explain the enlarged lattice parameter of the QD planes observed in the faceted steps (figure 4b). Slices through the volume are presented in a video as Supporting Information.

APT was performed in order to gain combined 3D structural and quantitative compositional information within the 3D extended defects. Figure 5b displays an APT Ga map of the tip-shape specimen obtained from sample S1. The Ga map enables the visualization of the QD planes, as they are those containing Ga. The upper part of the tip exhibits a homogeneous region with flat QD planes. However, a significant portion of the tip contains a threading defect corresponding to the faceted step present at the boundary of the cone-shaped extended defect. Figure 5c focuses on the APT analysis of 5 QD planes located on the lower part of the tip. The Ga-atom planes exhibit a bending angle of approximately 30°, which corresponds to the behavior observed in the BF-STEM image in figure 5d.

Figures 5e and 5f show projected side and top views of the APT Ga density map corresponding to the five QD layers in figure 5c, respectively. The side view is projected in-plane, capturing the thickness of the tip which is about 37 nm in diameter. The top view is projected along the growth direction, encompassing a thickness corresponding to 5 SL periods, which is approximately 27.5 nm. The QD layers exhibit a Ga enrichment towards the 30° facet. In order to quantify the chemical difference between the homogeneous region and the faceted area without artifacts caused by the preferential field evaporation[26], the quantification is done by the comparison of the Ga density, since the Al distribution appears homogenous along the specimen. Thus, considering that the non-defective regions, with a measured average Ga density of 1.0 atom/nm$^2$, contain $Al_{0.14}Ga_{0.86}N$ QDs (nominal concentration deduced from growth conditions), we deduce that the faceted region, with a measured average Ga density of 2.2 atom/nm$^2$, should contain $Al_{0.07}Ga_{0.93}N$ QDs, which should red shift the emission wavelength. This result is consistent with the findings from electron tomography, which indicated that the propagating structure presents Ga-rich boundaries. The variation in composition can be explained by the different mobility of Ga and Al adsorbed atoms. Al has very low mobility at the growth temperature and remains uniformly distributed upon deposition, whereas Ga atoms are highly mobile and can accumulate and incorporate at the steps due to Ehrlich-Schwoebel effect. In fact, preferential incorporation of Ga in similar steps has been reported in previous studies.[39,40]

The different chemistry of the dots can also explain the size variations observed: a higher metal-to-nitrogen ratio near the facets would lead to larger dots. However, the complex strain distribution at the facet (see figure 4b), due to the associated dislocation, might also have a significant impact in the SK growth.

To investigate whether the chemical and morphological changes of the QD population within the extended defects are responsible for the emission bimodality, we conducted 3D calculations of the electronic structure of AlGaN/AlN QDs by solving the Poisson and Schrödinger equations using Nextnano3, as described in the Experimental Methods.[33,34] Three structures were considered, taking into account different configurations of chemistry, morphology, and crystallography for the QDs:

a) The primary emission in the samples originates from QDs in defect-free regions. To simulate these QDs, we considered a QD superlattice oriented along the [0001] axis, with QDs containing 14% Al and the geometrical dimensions presented in figure 1d. According to our simulations, this structure produces an emission at 286 nm, which reproduces closely the experimental emission at 290 nm. Figure 6a depicts a representation of a single QD within the simulated structure, showing a {1-100} cross-section slice of the calculated electron and hole squared wavefunctions .

Then, in the defective region, we need to consider two different scenarios based on the experimental results obtained. Larger QDs were observed within the extended defects, and a Ga enrichment was found at the 30°-faceted boundaries. Therefore, two different simulations were performed:

b) QDs on the (0 0 0 1) plane containing 14% of Al and with a diameter of 7.2 nm, and a height of 6 ML. This simulation, with increased QD size, accounts for the QD population inside the extended defect. Figure 6b presents a {1-100} cross-section slice of the electron and hole wavefunctions in the QDs. The simulation predicts an emission at 298 nm, which closely matches the experimental value of 302 nm observed for the extended defects.

c) QDs on the 30° facet, containing only 7% of Al and with a diameter of 7.2 nm, and a height of 6 ML. Figure 6c presents a {1-100} cross-section view of this simulated structure. The electron and hole wavefunctions are separated both in-plane and along the growth axis, which reduces their overlap. This structure produces an emission at 299 nm, which matches the experimental value observed for the extended defects at 302 nm. The reduced component of the polarization perpendicular perpendicular to the AlGaN/GaN interfaces reduces the polarization-related red shift with respect to a), which partially compensates the effect of the reduced Al concentration. As a result, the emission wavelength is siminar to case b).

Therefore, our simulations conclude that the deformation of the QDs observed at the 30°-facets and within the propagated pits is responsible for the bimodal emission observed in this sample. The structural parameters used for the simulations and the corresponding results are summarized in table 1 and a scheme displaying the comparison of the experimental and theoretical results is presented in the figure 6d.

**Extension of the study to the 230 nm emission range**

The study presented above refers to sample S1 emitting at 294 nm, which was selected for this analysis for its compatibility with characterization using a Nd-YAG laser (266 nm). However, it is important to extend the study to the highly-demanded 230-270 nm range. LEDs emitting around 270 nm are starting to replace conventional mercury lamps,[6] and there is a growing interest in emitters around 230 nm since this wavelength poses no threat to human skin or eyes, making it a safe option for occupied spaces.[1,2] The potential utilization of this radiation for mitigating pandemics or combating antibiotic-resistant bacterial infections holds immense promise and could yield profound and transformative outcomes. However, this wavelength range poses additional challenges in terms of resistivity of the layers and light extraction issues: AlGaN material with the required Al content has a tendency to emit in-plane, and AlN presents an absorption band around 265 nm associated with pollutants. QDs are an interesting approach for far UV-C emitters because the 3D carrier confinement improves the radiative efficiency and the strain state of the dots favors the out-of-plane light extraction. However, the above-described bimodality of the QD emission becomes more important in samples emitting at wavelengths shorter than 270 nm.[13]

To extend our study to shorter wavelengths, we have considered sample S2, which contains 47% of Al in the QDs, with a main CL emission at 223 nm at 5 K and at 233 nm at room temperature, as shown in figure 7a. The emission wavelength of 233 nm at room temperature is ideal for disinfection applications as it strikes the optimal balance between maximum DNA absorption and penetration into the skin and cells for efficient and safe inactivation of bacteria and viruses.[2] Additionally, the estimated $IQE \approx 38\%$ is promising for the development of far UV-C emitters.[13] The main emission peak exhibits a FWHM of only 15 nm. However, the luminescence spectra display a clearly-resolved broad secondary emission at longer wavelengths, centered around 260 nm, both at room temperature and 5 K.

An SEM image of the S2 sample surface is displayed in figure 7b, revealing a pit density around $3 \times 10^8$ cm$^{-2}$, similar to S1. When comparing this morphology with the λ-filtered CL map shown in figure 7c (colors indicated in figure 7a), it becomes evident that the emission at longer wavelengths is associated with the pits, pointing to a similar behavior as that observed in S1. Here, the luminescence from the flat surface of the sample peaks around 233 nm, whereas the emission shifts to the 240-270 nm range in the vicinity of the pits. Notably, compared to the phenomenon observed in S1, the wavelength shift in S2 is significantly larger.

A combined structural and optical analysis was performed using TEM and CL on the same lamella specimen fabricated on the S2 sample. Figure 7d presents a BF-STEM image of the complete lamella, overlaid with a λ-filtered CL map (same color code as figure 7c). The figure reveals that the short wavelength emission originates from homogeneous regions of the sample, whereas the secondary emission at longer wavelengths originate from regions exhibiting contrast due to dislocations and associated surface roughening. The overall phenomenon appears to be, again, closely associated with dislocations arising from the AlN template.

Figure 7e displays a HAADF-STEM image of the interface between the AlN buffer layer and the AlGaN QD superlattice. Similar to S1, the formation of extended defects associated with dislocations is observed. However, in this sample, the 30° facets produce a defect structure with higher level of complexity. For instance, in the bottom-left part of the figure 7e, a pit, similar to those observed in the S1, is formed and propagates through the superlattice. However, the dislocation associated with the boundary of the extended defect nucleates a second extended defect several QD periods above the interface with the buffer layer. This is also observed at the top-left part of the image, where an extended defect arises from directly from a dislocation in the superlattice. The intermixing of these structures could explain the large FWHM of the secondary emission in S2. Regarding the larger red shift of the emission in S2 with respect to S1, it might be due to higher Ga enrichment in the defective areas of S2, as Ga represents a lower overall percentage of the QD chemical composition. This Ga enrichment could additionally contribute to the formation of larger QDs, offering further explanation for the observed increased redshift. In fact, the HAADF-STEM image presented in the figure 7e shows a high degree of chemical contrast between defective and non-defective regions, higher than that observed in figure 3b, which supports this hypothesis.

## 4. CONCLUSION

We have investigated the cause of the bimodal emission of $Al_xGa_{x-1}N/AlN$ QDs superlattices taylored for electron-beam pumped emitters, displaying high *IQE* in the 230-300 nm range. Our analysis has revealed that this bimodality is linked to the presence of cone-shaped extended defects originating at the first interface between the AlN buffer layer and the superlattice and propagating vertically. These defects are associated with a dislocation with a strong shear strain field, which motivates the generation of a pit with 30° faceted boundaries. We observe a Ga enrichment at the facets, which is attributed to the Ehrlich-Schwoebel effect, and an increased QD size within the defect domain. The different size and chemistry of the dots within the defect and at the defect boundary was identified as the source of the bimodality through the use of correlated microscopy techniques and 3D Schrödinger-Poisson calculations. Further investigations aimed at mitigating or reducing the occurrence of these structures will pave the way for the implementation of this efficient technology within disinfection applications. The use of bulk AlN substrates, with a significantly lower dislocation density, is a potential solution. However, finding cost-effective approaches might require creative thinking and alternative strategies.

■ **SUPPORTING INFORMATION**

Video showcasing the electron tomography study performed on a FIB-fabricated tip of the S1 sample.

■ **ACKNOWLEDGEMENTS**

This project received funding from the French National Research Agency (ANR) via the ASCESE-3D and FUSSL projects (ANR-21-CE50-0016 and ANR-22-CE09-0024). A CC-BY public copyright license has been applied by the authors to the present document and will be applied to all subsequent versions up to the Author Accepted Manuscript arising from this submission, in accordance with the grant's open access conditions.

Part of this work, performed on the Platform for NanoCharacterisation (PFNC) of CEA, was supported by the "Recherche Technologique de Base" Program of the French Ministry of Research.

The authors would also like to thank Matthew Bryan for improving the GEM_ED software by using Python and liberTEM libraries.

## ■ REFERENCES

Figure 1

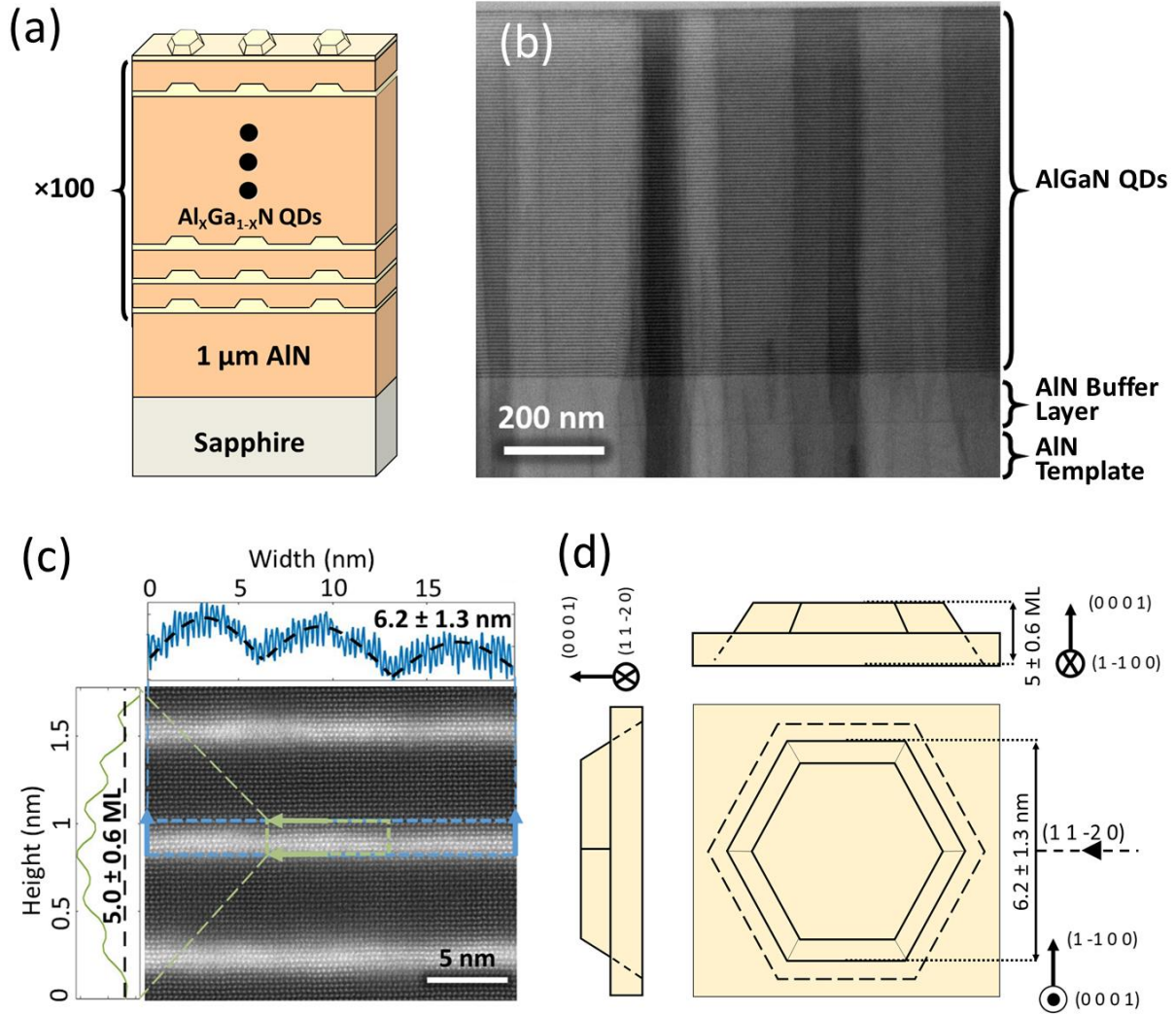

**Figure 1.** (a) Sample structure consisting of 100 periods of $Al_xGa_{1-x}N$ QDs embedded in AlN barriers. (b) BF-STEM image of the sample S1 along the (1 1 -2 0) axis showing the AlN template, the AlN buffer layer, and the AlGaN/AlN QD superlattice. (c) HAADF-STEM zoomed-in image showing three QD planes in the middle of the SL and the measurement of QD widths and heights through intensity profiles measured on the dashed rectangles. (d) QD schematic front, lateral, and top views; the sizes are deduced from the measurements shown in (c).

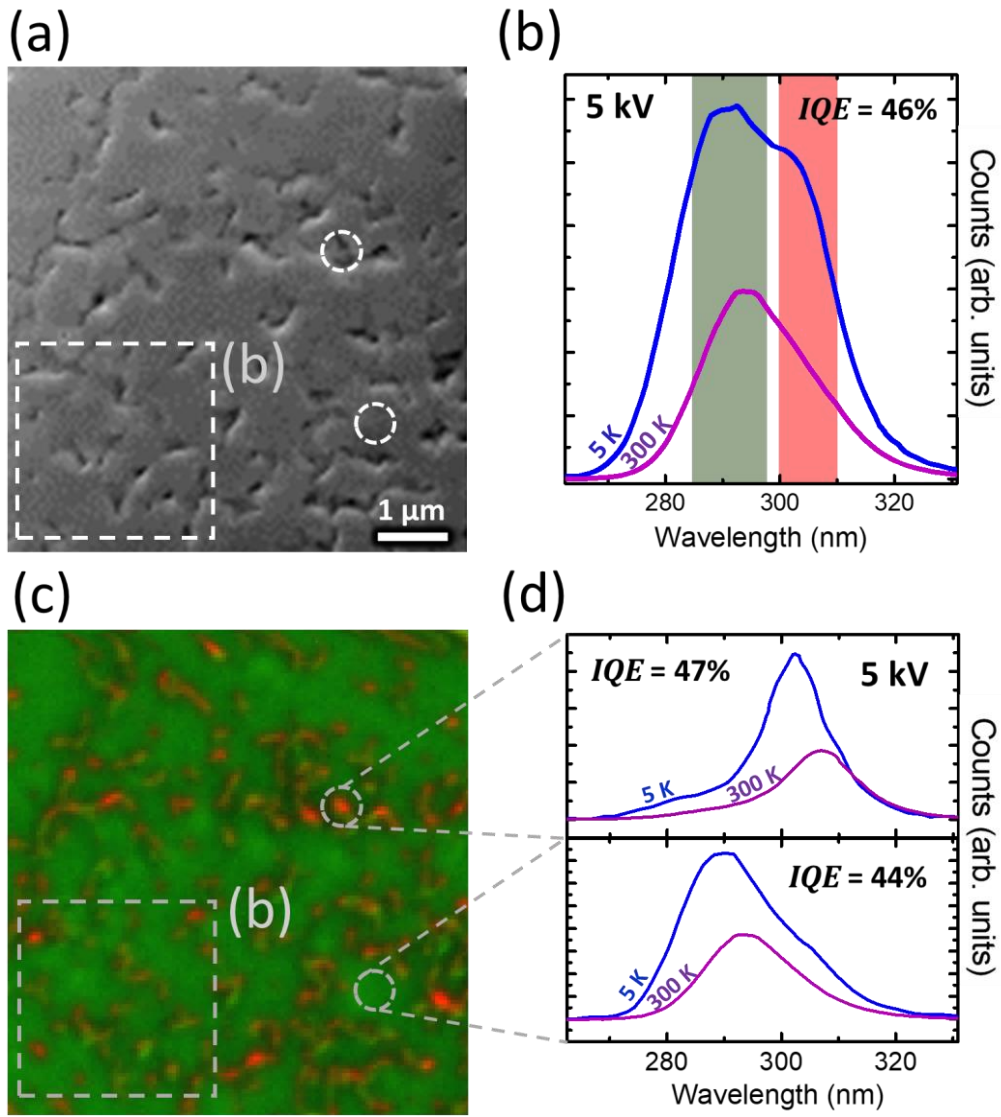

**Figure 2.** (a) Top-view SEM image of the S1 sample. (b) 5kV CL emission spectra from the S1 sample surface acquired on the dashed square region of (a) at 300 K and 5 K. (c) λ-filtered CL map of the sample at room temperature, covering the same region as (a), with the color code corresponding to the wavelengths shown in color in (b). (d) Local CL emission spectra from the vicinity of the pits (top) and from the flat surface (bottom).

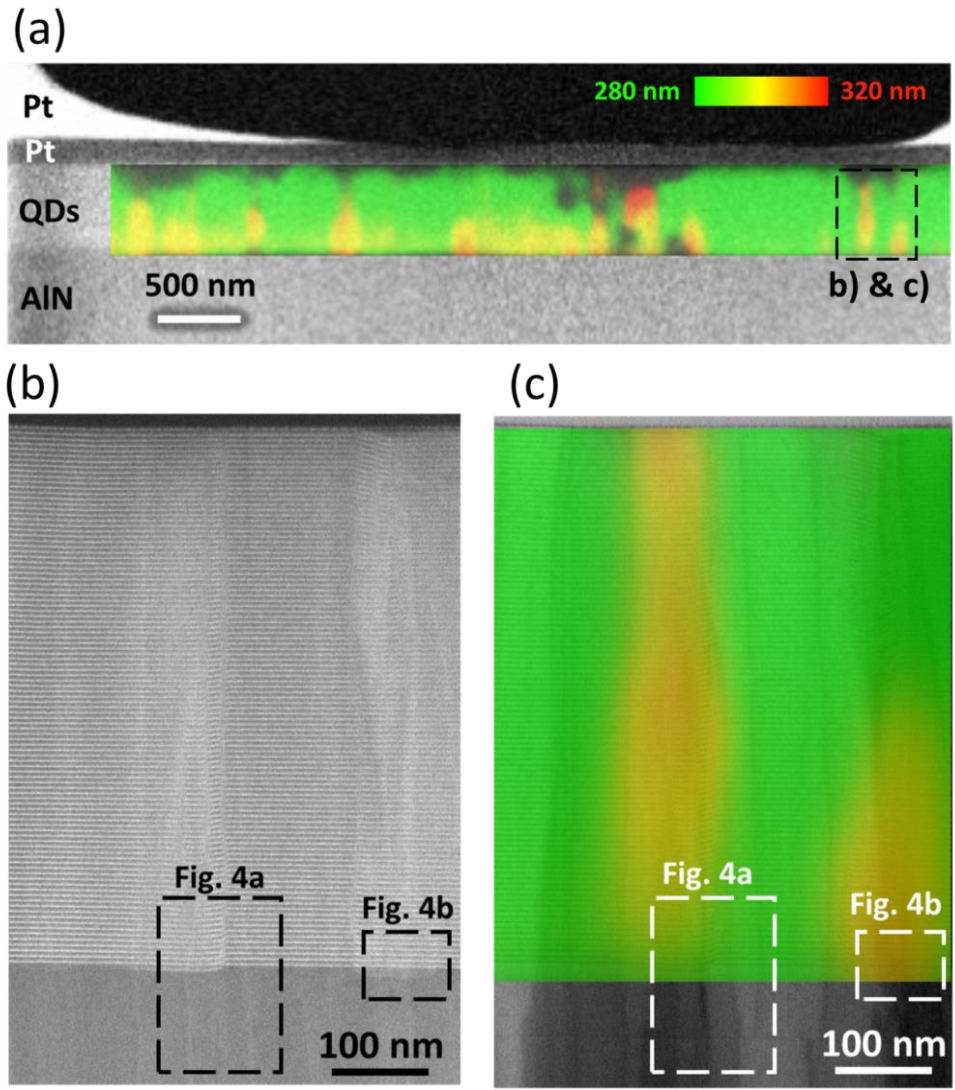

**Figure 3.** (a) SEM image of a FIB-lamella from the sample with a superimposed λ-filtered CL map. (b) HAADF-STEM image from the dashed rectangle shown in (a) showing two cone-shaped extended defects. (c) BF-STEM image from the same region as (b) with a superimposed λ-filtered CL map.

**Figure 4**

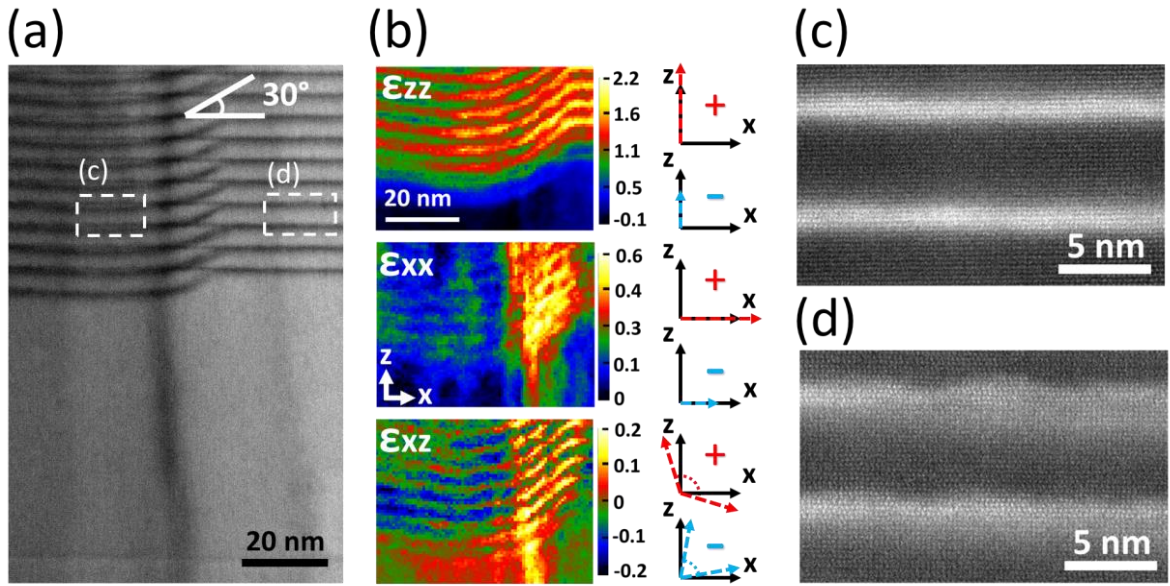

**Figure 4.** (a) STEM-BF image acquired on the origin of an extended defect, as depicted on the left dashed rectangle in figure 3b. (b) Deformation maps ($\varepsilon_{zz}$, $\varepsilon_{xx}$ and $\varepsilon_{xz}$) of the origin of an extended defect, as shown on the right dashed rectangle in figure 3b. These maps include the dislocation and the faceted step formed at the interface between the AlN buffer layer and the AlGaN/AlN QD superlattice. The arrow schemes on the right side of the maps illustrate the meaning of positive and negative deformation in each image. (c-d) HAADF-STEM zoomed-in images from two QD planes from (c) outside and (d) inside the extended defect, as indicated in (a).

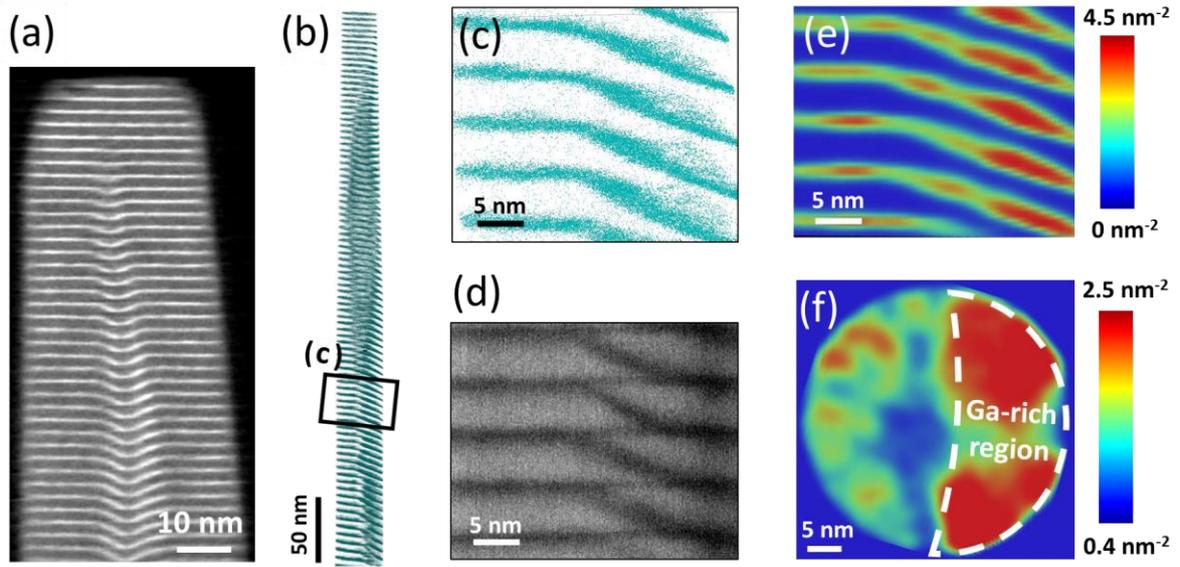

**Figure 5.** (a) Electron tomography slice of a tip prepared on the S1 sample containing a pit that propagates through the superlattice generating a cone-shaped extended defect. (b) Ga APT reconstruction for a different tip prepared on the S1 sample. (c) Ga APT reconstruction, (d) STEM-BF image and (e) Ga APT projected density map extracted from five QD planes located at the boundary of an extended defect. The region corresponds to the zone highlighted in the square in figure 5b. (f) Top-view Ga APT projected density map from the same five QD planes as (e). The projected density maps units are arbitrary.

**Figure 6**

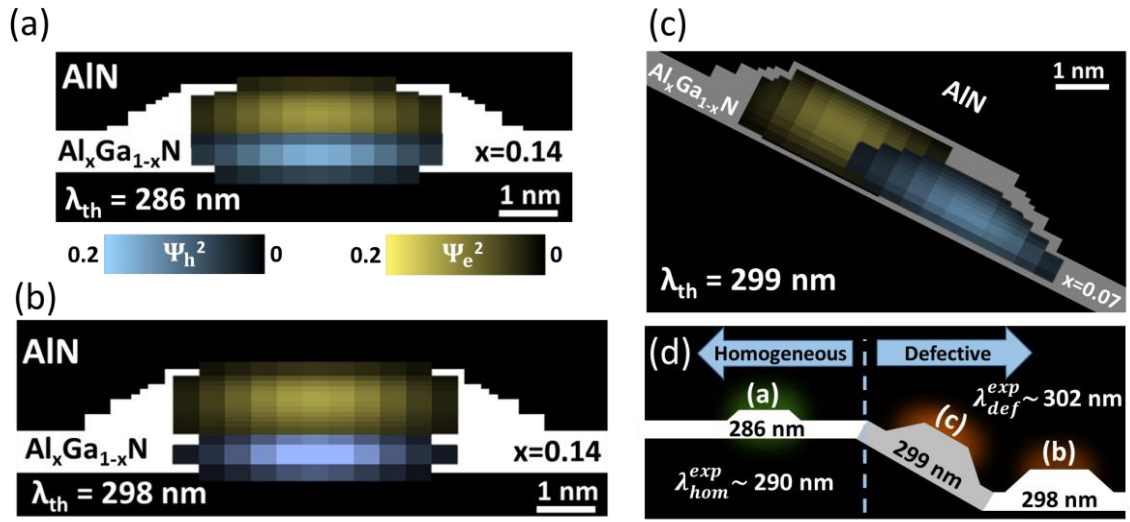

**Figure 6.** Nextnano3 simulations of the electron and hole wavefunctions, and expected emission wavelength in structures reproducing the QD morphology and chemistry for different regions of the sample: (a) QDs from the homogeneous region, (b) QDs from interior of the defective region, and (c) QDs located at the Ga-rich facets. The diameter, height and composition of the QDs and the results from the simulation are detailed in table 1. (d) Scheme of the simulated QD structures and their respective extracted wavelength compared to the experimental value.

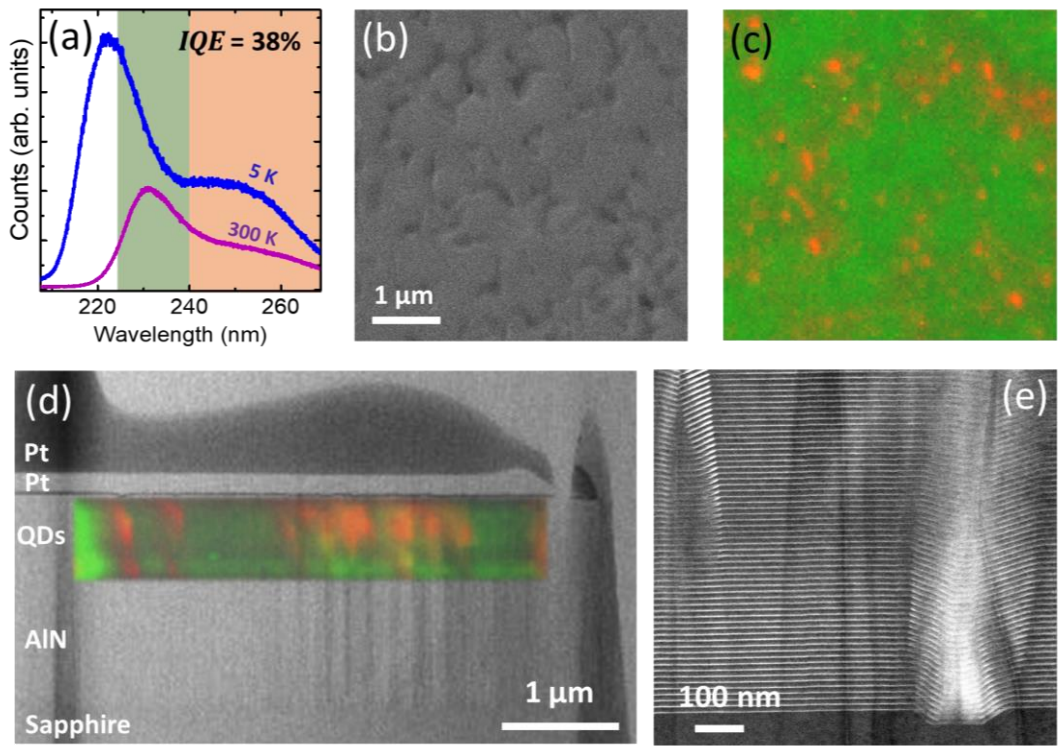

**Figure 7.** (a) CL emission spectra from the S2 sample surface at 300 K and 5 K. (b) Top-view SEM image and (c) λ-filtered CL map of the sample S2 at RT. The color code correspond to the wavelengths shown in color in (a). (d) BF-STEM image of a FIB-lamella from the sample S2 with a superimposed λ-filtered CL map. (e) HAADF-STEM image of the interface between the AlN buffer layer and the AlGaN QD superlattice showing the nucleation of cone-shaped extended defects.

# Tables

| Region | h (ML) | D (nm) | Al (%) | z axis | $E_e$ (eV) | $E_h$ (eV) | $E_{ex}$ (eV) | $\lambda$ (nm) |
|---|---|---|---|---|---|---|---|---|
| (a) | 5 | 6.2 | 14 | [0 0 0 1] | 4.06 | -0.33 | 0.06 | 286.0 |
| (b) | 6 | 7.2 | 7 | [5 -5 0 9.2] | 0.37 | -3.84 | 0.06 | 298.7 |
| (c) | 6 | 7.2 | 14 | [0 0 0 1] | 3.85 | -0.36 | 0.05 | 297.8 |

**Table 1.** Input parameters for the Nextnano simulation of QDs in the three considered regions (as defined in figure 6d), including QD heigth (h), diameter (D), Al atomic percentage, and the normal vector to the QD plane (z axis). Results of the calculations, including first electron energy level ($E_e$), first hole energy level ($E_h$), exciton binding energy ($E_{ex}$), and estimated emission wavelength ($\lambda$).